\documentclass[12pt,preprint]{aastex}
\begin{document}

\def\be{\begin{equation}}
\def\ee{\end{equation}}
\def\lesssim{\raisebox{-0.3ex}{\mbox{$\stackrel{<}{_\sim} \,$}}}
\def\gtrsim{\raisebox{-0.3ex}{\mbox{$\stackrel{>}{_\sim} \,$}}}

\title{Using the Intensity Modulation Index to Test Pulsar Radio Emission Models}

\author{Fredrick A. Jenet \altaffilmark{1}, Janusz Gil\altaffilmark{2}}
\altaffiltext{1}{California Institute of Technology, Jet Propulsion Laboratory\\ 4800 Oak Grove Drive,Pasadena, CA 91109}
\altaffiltext{2}{Institute of Astronomy, University of Zielona G\'ora \\Lubuska 2, 65-265, Zielona G\'ora, Poland}

\begin{abstract}
This letter explores the possibility of testing pulsar radio emission
models by observing pulse-to-pulse intensity modulation. It is shown
that a relationship between a pulsar's period, period derivative, and
intensity modulation is a natural consequence of at least one
theoretical model of radio pulsar emission. It is proposed that other
models may also predict a similar correlation. The exact form of the
relationship will depend on the model in question. Hence, observations
of intensity modulation should be able to determine the validity of
the various emission models. In an attempt to search for the predicted
dependencies, the modulation properties of a set of 12 pulsars are
studied. These data are suggestive, but they are unable to
differentiate between three possibilities for the emission
process. Future observations will be able to confirm these results and
determine whether or not specific emission models are viable.

\end{abstract}


\section{Introduction}
The cause of the emission from radio pulsars has remained elusive
since their discovery over 30 years ago. The high brightness
temperature together with the enormous amount of phenomenology
exhibited make these sources very difficult to understand. This letter
focuses on pulse-to-pulse fluctuations and it will describe how they
may be used to test emission models.

Observations of bright pulsars have shown that the shapes and
intensities of individual pulses are unique, although they average
together to form a stable mean profile. The characteristic widths of
individual pulses, typically referred to as sub-pulses, are usually
smaller than the average profile width. Some pulsars show rapid
intensity fluctuations or micro-structure. The time scales of these
fluctuations vary from source to source and they range from 1 ms down
to 2 ns \citep{hkw+03}. 



Recent observations of PSR B1937+21 show a behavior that is
completely different from previously studied sources \citep[][J01
hereafter]{jetal01}. This source exhibits no detectable pulse-to-pulse
fluctuations. Occasional bursts of radio radiation, or ``Giant
pulses'', are observed but they are restricted to small regions in
pulse phase \citep{kt00, cst+96}. Understanding what makes the
non-giant pulse emission of PSR B1937+21 so unique will help us to
understand the radio emission process. The possibility that this
steady behavior is just an extreme case of a general phenomenon is explored
in this letter.



The ideas presented here will be focused on the modulation index,
which is a measure of pulse-to-pulse intensity fluctuations. The
modulation index is known to be a function of pulsar pulse phase, hence
one may define a phase resolved modulation index as follows:
\begin{equation}
m(\phi) =\frac{ \sqrt{<I(\phi)^2> -<I(\phi)>^2}}{<I(\phi)>},
\end{equation}
where $m$ is the modulation index, $\phi$ is the pulse phase, $I$ is
the pulsar intensity, and the angle brackets represent averaging over
a large ensemble of adjacent pulses. Recent theoretical work by
\citet[][GS00 henceforth]{gs00} suggests that the pulsar intensity
modulation index should depend on some function of the pulsar period
($P$) and period derivative ($\dot{P}$). The exact functional
dependence will depend on the region of pulse phase being
studied. More specifically, it will depend on whether the phase region
is classified as a ``core'' or ``conal'' component as defined by
\citet{r83,r86}. Due to current constraints on the available data and
the predictions of the GS00 model, the work here will focus on core
component emission only.

The GS00 model is based on the polar cap spark model of
\citet{rs75}. Both of these models have received much attention in
recent years. They have been used to interpret the sub-pulse
properties of slow pulsar conal emission
\citep{esv03,rr03,vsr+03,ad01,dr01} and of millisecond pulsar core
emission \citep{es03}. These models have also been used in pulsar
population studies \citep{acc02,fcm01}. 

In general, a given theory of the emission physics should be able to
make quantitative predictions for the dependence of $m$ on $P$ and
$\dot{P}$. Observations should then be able to rule out various
classes of models. The predictions of the GS00 model together with other
possible models are discussed in section 2. Section 3 demonstrates how
these models can be tested. This letter is summarized in section 4.

\section{Models of Intensity Modulation}

It is generally accepted that the observed pulsar radio emission is
generated within a dense electron-positron plasma flowing along the
open magnetic field lines of the neutron star. Open field lines are
those that connect with the interstellar magnetic field rather than
connecting back to the pulsar's surface magnetic field. Intrinsic
pulse-to-pulse intensity modulations can arise from the time-dependent
lateral structure of this flow, probed once per pulsar period by the
observer's line-of-sight. 

\citet{rs75} proposed a pulsar model where bursts of plasma, or
``sparks'', are generated at the polar cap. The electron-positron
plasma created by a spark travels up along the magnetic field lines
where it eventually emits radio radiation generated by some kind of
instability \citep{am98,mgp00}. GS00 explored the Ruderman \&
Sutherland model further in an attempt to describe pulsar radiation
properties as a function of basic observable parameters such as $P$
and $\dot{P}$. They postulated that the polar cap is populated as
densely as possible with a number of these sparks, each having a
characteristic size and separation from adjacent sparks that is
approximately equal to the gap height $h$. This leads directly to the
so called ``complexity parameter'' $a_1=r_p/h$, equal to the ratio of
the polar cap radius, $r_p$, to the characteristic spark dimension,
$h$. Making a reasonable assumption about the non-dipolar surface
magnetic field, GS00 found that \be
a_1=5(\dot{P}/10^{-15})^{2/7}(P/1s)^{-9/14} .\label{a} \ee 
One can show that $a_1$ is
the maximum number of sparks across the polar cap. It is also the
maximum number of subpulses and/or profile components. Thus, $a_1$
describes the complexity of the mean pulse profile (see GS00 for
details). Within a given polar cap region (i.e. core or conal
region), the amplitude of the emitted radio radiation is roughly the
same for each spark. Since each individual spark emits nearly steady,
unmodulated radiation, the observed pulse-to-pulse fluctuations are
due to the presence of several sparks moving either erratically or in
an organized manner and emitting into the observers line of sight. As
the number of sparks increases, one will expect to see less and less
pulse-to-pulse intensity fluctuation. Hence, the modulation index
should be anti-correlated to the complexity parameter in both core and
conal components. Unfortunately, other effects such as those
associated with viewing angle will mask this anti-correlation in the
conal emission. Thus, the core emission is the most direct way to
observe this effect. 




The GS00 model is based on instabilities in the polar cap plasma
generation. There are several other magnetospheric instabilities that
could, in principle, produce something like a complexity parameter
that would be correlated to the modulation index. Three such
instabilities are: continuous current outflow instabilities
\citep{as79,ha01}, surface magnetohydrodynamic wave instabilities
\citep{lou01} and possibly outer magnetospheric instabilities. Even
though a complexity parameter has not been rigorously calculated
for these models, one can estimate that the following parameters:
\begin{equation}
a_2 = \sqrt{\frac{\dot{P}}{P^3}},~~a_3 = \sqrt{P \dot{P}},~~a_4 = \sqrt{\frac{\dot{P}}{P^5}},
\end{equation}
would correspond to the complexity-like parameters for the current
outflow, surface MHD wave, and outer magnetospheric instabilities,
respectively. Physically, these parameters are proportional to the
total current outflow from the polar cap, the surface magnetic field,
and the magnetic field at the light-cylinder, respectively.





\section{Analyzing the Intensity Modulation Properties}

A comparison between the observed modulation indices of 12 pulsars and
the various complexity parameters defined above is performed in this
section. Data were obtained for 8 sources from \citet[][W86
hereafter]{wetal86}, 2 from J01, and 2 from recent data taken at the
Arecibo Observatory using the Caltech Baseband Recorder. These sources
are listed in Table \ref{table1} along with the measured modulation
indices, observing frequencies, and references.

This study is focused on the emission properties of core components
since the current form of the GS00 model is more directly applicable
to core type emission. In general, the modulation indices of
core type emission are lower than that of conal emission (W86). This
effect is also a consequence of the GS00 model. For the case of
multiple component profiles, if conal emission overlaps with core
emission, the observed modulation index will be larger than that of
the core emission alone. Even pulsars that are classified as primarily
core emitters can have some conal emission near the edges of the
profile (see $\S$ 5.4 in GS00). In order to reduce the effects of
overlapping emission regions, the minimum value of the modulation index
was chosen for each source. This will result in the best possible
measurement of the core component's modulation index.

 
Since the data from J01 were reported using a definition of
the modulation index that included radiometer noise, the values were
transformed in order to be consistent with the definition in W86. The
following transformation was applied (see J01 for details):
\begin{equation}
\label{mxform}
m = \sqrt{(m_{j}^2 -1)/2},
\end{equation}
where $m$ is the modulation index used in W86 as well as in this
letter and $m_j$ is the modulation index used in J01.

The measured modulation index depends on both intrinsic pulsar intensity fluctuations as well as fluctuations due
to propagation through the interstellar medium (ISM). The functional form
of this dependence is as follows:
\begin{equation}
m^2 + 1 = (m_{i}^2 + 1)(m_{ISM}^2 +1),
\label{mod_conv}
\end{equation}
where $m$, $m_i$ and $m_{ISM}$ are the measured, intrinsic, and ISM
induced modulation indices, respectively. $m_{ISM}$ may be estimated
using the following relationship \citep{cwd+90}:
\begin{equation}
\label{mism}
m_{ISM} = 1/\sqrt{S},
\end{equation}
where $S$ is the number of ``scintills'' in the receiver bandwidth. $S$ is given by
\begin{equation}
\label{sism}
S = 1 + \eta \frac{B}{\delta \nu},
\end{equation}
where $B$ is the receiver bandwidth, $\delta \nu$ is the ISM
decorrelation bandwidth, and $\eta$ is a filling factor which ranges
from 0.1 to 0.2. For each source, $\delta \nu$ was taken from
\citet{cor86}, $\eta$ was set to 0.18, and the intrinsic modulation
index was estimated using Equation \ref{mod_conv}. Note that the
results presented here are insensitive to variations in $\eta$ when
this parameter is within the expected range. The adopted value of
$\eta$ was chosen so that each inferred intrinsic modulation index was
non-zero.

Three criteria were used to select the sources used in this
study. First, a given source had to have a measured period derivative
\citep{tml93}. Second, the ISM decorrelation bandwidth must be known
\citep{cor86}. Third, the source had to have a core emission
component.

In order to determine if any of the complexity parameters are
correlated with the measured modulation indices, the Spearman
Rank-Order Correlation (SROC) coefficient, $\rho$, and its associated
significance parameter, $\Delta$, are calculated between $m$ and each
$a_i$.  $\Delta$ is simply the probability that such a correlation
would occur in randomly distributed data. Hence, the smaller the value
of $\Delta$, the more significant the correlation. The SROC
coefficient was chosen over other possible statistics for two
reasons. First, it is more robust and conservative then the standard
linear correlation coefficient (see $\S 14.6$ of Press et
al. (1992)). Second, since it is a rank ordering method, $|\rho|$ and
$\Delta$ are exactly the same for both the original data,($x_j,y_j$),
and for ($F(x_j),G(y_j)$) where $F$ and $G$ are arbitrary, monotonic
functions. This property is extremely useful since the current form of
the GS00 model only predicts the existence of a relationship between
$m$ and $a_1$ rather than specifying an exact form.

Plots of $m$ versus $a_1$, $a_2$, $a_3$, and $a_4$ are shown in Figure
1. The error bars shown were taken to be the greater of the
measurement uncertainty or the uncertainty due to the fact that $\eta$
is unknown and can range from 0.1 to 0.2. The SROC coefficient,
$\rho$, and the significance parameter, $\Delta$, were calculated for the data,
given each of the models. The results are tabulated in Table
\ref{table2}. $\rho$ and $\Delta$ were calculated both with and
without the ISM correction applied. For the sake of comparison, the
correlation of $m$ with $P$ and with $\dot{P}$ were calculated and
included in the table. The sparking gap model $a_1$ of GS00 shows the
best correlation, although $a_2$ and $a_4$ cannot be excluded.

The above analysis calculated the correlation between the intensity
modulation index and four physically motivated parameters. An
alternative to this approach is to calculate the correlation between
$m$ and a set of parameters given by the following one-dimensional family:
\begin{equation}
a(\alpha) = P^\alpha \dot{P}.
\label{fam1}
\end{equation}
One can then find that $\alpha$ which maximizes both the absolute
value of the correlation and its significance. This is equivalent to
minimizing the significance parameter since $\Delta$ is a monotonic
function of $|\rho|$. A range of admittable $\alpha$ values about this
minimum can be obtained by choosing a threshold value of $\Delta$.
Since the SROC analysis is independent of an arbitrary monotonic
function, it is not necessary to search over the two dimensional
family of the form $P^\beta \dot{P}^\gamma$. For the data presented in
Table \ref{table1} together with a threshold significance parameter of
$1 \times 10^{-3}$, $\alpha$ ranged from $-5.0$ to $-2.0$ with a local
minimum located at $-2.7$. The minimum $\Delta$ was $2.8 \times
10^{-5}$ and the corresponding correlation coefficient was
$-0.92$. The range of $\alpha$ values searched over was [$-$100,100]
with a grid spacing of 0.01. $\Delta(\alpha)$
varies in a piece-wise continuous manner with only one local minimum
that is also the global minimum in the region searched. 

In order to determine the significance of the value of $\alpha$ found
using the method described above, Monte-Carlo techniques were used to
determine the probability of obtaining an $\alpha$ with $\Delta \leq
2.8 \times 10^{-5}$. For the same set of pulsars used above, random
modulation indices were calculated and the minimum significance was
found over a range of $\alpha$ values equal to [$-$20,20]. Note that
this range is smaller than that used above in order to reduce
computation time. The grid spacing used here was also 0.01. When the
modulation indices are chosen from a uniform distribution ranging from
0 to 1, the probability of obtaining $\Delta \leq 2.8 \times
10^{-5}$ is $.0011 \pm 9\%$. A random set of $m$ values may also be
obtained by randomly shuffling the measured set of modulation
indices. When this is done, the probability becomes $.00074 \pm 12\%$.

\begin{figure}
\plotone{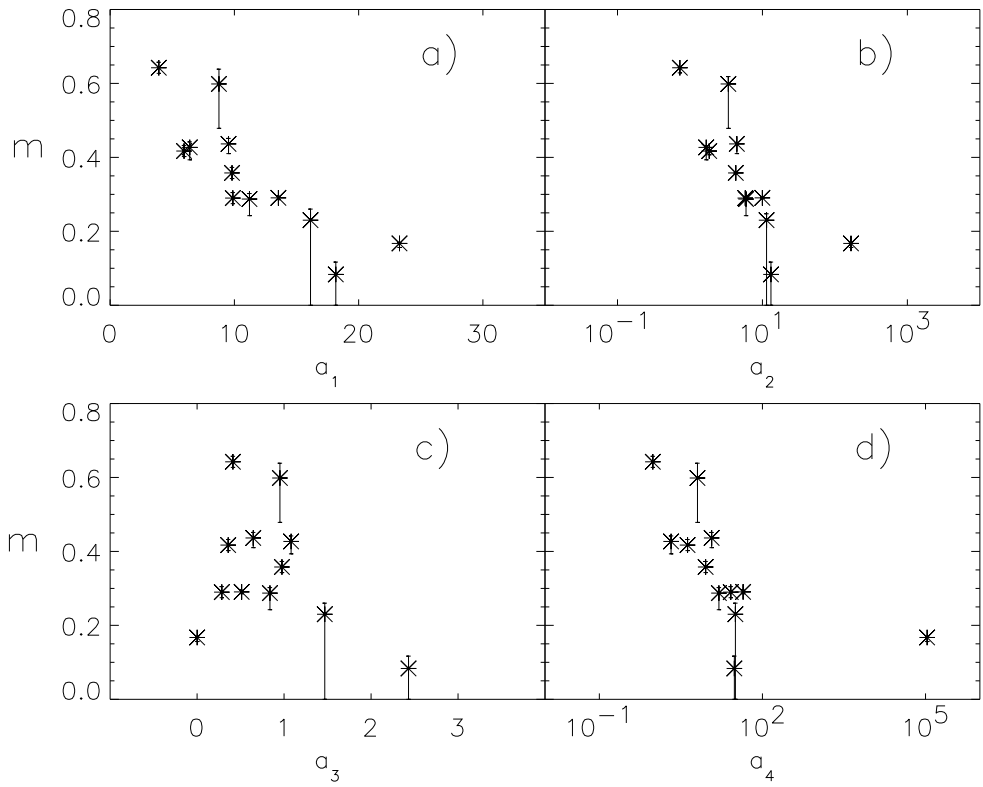}

\caption{$m$ versus different ``complexity parameters''. These parameters are defined in Table \ref{table2}. A log scale was used when the parameter varied over several orders of magnitude.}

\label{figure1}
\end{figure}

\begin{deluxetable}{ccccc}
\tablecaption{\label{table1}Sources used in this analysis}
\tabletypesize{\scriptsize}
\tablewidth{3in}

\tablehead{\colhead{Source} & \colhead{m} & \colhead{Frequency} & \colhead{Reference} \\ \colhead{(B1950)} & &\colhead{(MHz)} &}

\startdata
0626+24 & \mbox{$0.36\pm 0.015$} & 430 & W\\
0823+26 & \mbox{$0.88\pm 0.04$} & 430 & J\\
0919+06 & \mbox{$0.29\pm 0.015$} & 430 & W\\
1737+13 & \mbox{$0.47\pm 0.015$} & 430 & W\\
1821+05 & \mbox{$0.66\pm 0.015$} & 430 & W\\
1839+09 & \mbox{$0.47\pm 0.015$} & 430 & W\\
1842+14 & \mbox{$0.34\pm 0.015$} & 430 & W\\
1929+10 & \mbox{$1.08\pm 0.005$} & 1410 & N\\
1933+16 & \mbox{$0.38\pm 0.03$} & 1666 & N\\
2053+36 & \mbox{$0.29\pm 0.015$} & 430 & W\\
2113+14 & \mbox{$0.43\pm 0.015$} & 430 & W\\
1937+21 & \mbox{$0.17\pm0.003$}  & 430 & J\\

\enddata

\tablerefs{W = \citep{wetal86}, J=\citep{jetal01}, N=this paper}

\end{deluxetable}

\begin{deluxetable}{ccccc}
\tablewidth{5in}
\tablecaption{\label{table2}The Spearman rank-ordered correlation coefficients for several emission models}

\tablehead{\colhead{{\scriptsize Parameter}} & \colhead{{\scriptsize Model Type}} & \colhead {\scriptsize Definition} &\colhead{{\scriptsize \mbox{$\rho(\Delta)$}}} & \colhead{{\scriptsize \mbox{$\rho(\Delta)$}}} \\ & & & & {\scriptsize w/o ISM } \\ & & & & {\scriptsize correction}}
\tabletypesize{\scriptsize}
\startdata
{\scriptsize \mbox{$a_1$}} & {\scriptsize Sparking Gap }& {\scriptsize \mbox{$5(\dot{P})^{2/7}P^{-9/14}$}}& 
{\scriptsize $-0.91$($4\times 10^{-5}$) }&  {\scriptsize $-0.59(0.04)$}\\
{\scriptsize \mbox{$a_2$}} & {\scriptsize Beam} &{\scriptsize \mbox{$\dot{P}^{1/2}P^{-3/2}$}}&{\scriptsize $-0.90$($1\times 10^{-4}$) }& {\scriptsize $-0.58(0.05)$} \\
& {\scriptsize Instability} & & & \\
{\scriptsize \mbox{$a_3$}} & {\scriptsize MHD Waves } &{\scriptsize \mbox{$\dot{P}^{1/2}P^{1/2}$}} &{\scriptsize $-0.13(0.68)$} & {\scriptsize $0.09(0.78)$} \\
{\scriptsize \mbox{$a_4$}} & {\scriptsize Instabilities} & {\scriptsize \mbox{$\dot{P}^{1/2}P^{-5/2}$}} &{\scriptsize $-0.82$($1\times 10^{-3}$) }& {\scriptsize $-0.45(0.14)$} \\
& {\scriptsize at LC }& & & \\
& & {\scriptsize \mbox{$P$}} & {\scriptsize $0.69(0.01)$} & {\scriptsize $0.49(0.10)$} \\
& & {\scriptsize \mbox{$\dot{P}$}} &{\scriptsize $-0.34(0.28)$} &{\scriptsize $-0.08(0.81)$} \\
\enddata

\tablecomments{The Spearman Rank-Ordered Correlation coefficient,
$\rho$, and its significance, $\Delta$, are calculated between the
modulation index, $m$, and the complexity parameters associated with
four different emission models. For the sake of comparison, the table
also lists the correlation between $m$ and both $P$ and $\dot{P}$. For
all cases, $P$ is in seconds and $\dot{P}$ is in $10^{-15}s/s$. The
correlations were calculated both with and without the intersteller
medium correction applied. $\Delta$ is the probability of
obtaining this correlation in random data.}
\end{deluxetable}

\clearpage

\section{Discussion \& Summary}

The set of 12 pulsars studied here suggests a relationship between the
intensity modulation index, the pulsar period, and the period
derivative. Future observations are needed in order to confirm this
correlation. The search for such a relationship is an extremely
powerful way to constrain emission mechanisms. Using ``reasonable''
assumptions about the pulsar magnetospheric plasma, the sparking gap
model predicts a functional relationship between the modulation index,
the period, and the period derivative. If this correlation is shown
not to exist, then the assumptions, and perhaps the entire model, are
incorrect. The same may hold true for the other models discussed here,
if it can be shown that such correlations should exist. On the other
hand, if the correlation seen here is confirmed, then the exact
functional relationship will be able to determine which model, if any,
is the most likely candidate. The current data supports complexity
parameters of the form given by Equation \ref{fam1} with $\alpha$
between $-5.0$ and $-2.0$. Among the physical models presented, the
sparking gap model, $a_1$ ($\alpha = -2.25$), shows the highest
correlation, although the beam current model, $a_2$ ($\alpha = -3.0$), and
the light cylinder model, $a_4$ ($\alpha = -5.0$), cannot be
ruled out. The surface MHD wave model, $a_3$ ($\alpha = 1.0$), is
unlikely. The minimization analysis favors $\alpha=-2.7$ but the
corresponding values of $\rho$ and $\Delta$ are only slightly better
then those found for the sparking gap model. If follow-up observations
confirm that $\alpha=-2.7$, then none of the above models fully
capture the physics of the emission process.

Future observations will provide a data set far superior to the one
used in this analysis. Using the statistical techniques employed by
J01, the modulation indices of a much larger sample of pulsars can be
measured. Also, the ISM parameters, $\eta$ and $\delta \nu$, which are
known to vary with time, can be measured simultaneously with the
modulation index. This will enable a more accurate determination of
the intrinsic modulation index. Note that for this work, $\eta$ was
assumed to be $0.18$ for all sources and the decorrelation bandwidths
were taken from previously published results.





It should be noted that recent work on the Vela pulsar shows that this
source exhibits large pulse-to-pulse modulation \citep{kjs02}. In each
of the supported models, this source should have almost no
modulation. Since this pulsar is classified as a core emitter, it will
be an exception to the work presented here. If future observations
confirm the above correlation, then this source may be understood within the
framework of the supported models. For example, in the context of the
sparking gap model this pulsar may have surface magnetic field
structures or relativistic plasma $\gamma$ factors which differ
significantly from the main group. It is also possible that Vela may belong
to a class of pulsars which obey a different $m$, $P$, and $\dot{P}$ relationship.


The ideas presented here were motivated by observations of PSR
B1937+21 in which no detectable pulse-to-pulse modulation was
found. In the context of each of the models discussed above, the
stability of this pulsar's emission would be a consequence of its
relatively high complexity parameter. The physical reason for the
stability will be constrained further when future observations confirm
the correlation discussed here and determine its functional form more
accurately.

In summary, the relationship between a pulsar's pulse-to-pulse
intensity fluctuations, period, and period derivative will provide a
valuable insight into the physical processes responsible for the radio
emission. Such a relationship could offer a simple explanation for
the unique behavior observed in PSR B1937+21. The data presented in
this letter supports such a relationship although future observations
are needed in order to confirm its existence.

\acknowledgements 

This paper is supported in part by the grant 2 P03D 008 19 of the
Polish State Committee for scientific research. Part of this research
was performed at the Jet Propulsion Laboratory, California Institute
of Technology, under contract with the National Aeronautics and Space
Administration. The authors wish to thank John Armstrong and Linqing
Wen for useful discussions, Stuart Anderson for helping to take the
data, and the anonymous referee who made several useful
suggestions. Special thanks goes to E. B. Dussan V. and Thomas
A. Prince.

{}
\end{document}